\documentclass[letterpaper, 10 pt, conference]{ieeeconf}  % Comment this line out
\IEEEoverridecommandlockouts                              % This command is only
\usepackage{graphicx,amssymb,amstext}
\usepackage{amsmath}
\usepackage{tikz,lscape}
\usepackage{hyperref}
\usepackage{epsfig}
\usetikzlibrary{shapes,arrows}
\usepackage{verbatim}

\title{\LARGE \bf
Robust Filtering for Adaptive Homodyne Estimation of\\ 
Continuously Varying Optical Phase
}

\author{Shibdas Roy$^{1}$*, Ian R. Petersen$^{2}$ and Elanor H. Huntington$^{3}$%
\thanks{$^{1}$S. Roy, $^{2}$I. R. Petersen and $^{3}$E. H. Huntington are with the School of Engineering and Information Technology, University of New South Wales, Canberra.}%
\thanks{*\tt\small shibdas.roy at student.adfa.edu.au}%
}

\begin{document}

\tikzstyle{block} = [draw, fill=white, rectangle, 
    minimum height=2em, minimum width=4em]
\tikzstyle{sum} = [draw, fill=white, circle, node distance=1cm]
\tikzstyle{open}=[inner sep=1mm]
\tikzstyle{none}=[coordinate]
\tikzstyle{pinstyle} = [pin edge={to-,thin,black}]
\tikzstyle{pt}=[regular polygon,regular polygon sides=3,regular polygon rotate=-90, draw=black,scale=0.7,inner sep=-0.5pt]

\maketitle
\thispagestyle{empty}
\pagestyle{empty}

%%%%%%%%%%%%%%%%%%%%%%%%%%%%%%%%%%%%%%%%%%%%%%%%%%%%%%%%%%%%%%%%%%%%%%%%%%%%%%%%
\begin{abstract}

Recently, it has been demonstrated experimentally that adaptive estimation of a continuously varying optical phase provides superior accuracy in the phase estimate compared to static estimation. Here, we show that the mean-square error in the adaptive phase estimate may be further reduced for the stochastic noise process considered by using an optimal Kalman filter in the feedback loop. Further, the estimation process can be made robust to fluctuations in the underlying parameters of the noise process modulating the system phase to be estimated. This has been done using a guaranteed cost robust filter.

\end{abstract}

%%%%%%%%%%%%%%%%%%%%%%%%%%%%%%%%%%%%%%%%%%%%%%%%%%%%%%%%%%%%%%%%%%%%%%%%%%%%%%%%
\section{INTRODUCTION}

\bstctlcite{BSTcontrol}

Quantum parameter estimation is the problem of estimating an unknown classical parameter, often an optical phase shift, of a quantum system \cite{WM}. It is at the heart of many fields such as gravitational wave interferometry \cite{GMM}, quantum computing \cite{HWA} and quantum key distribution \cite{IWY}. The fundamental limit to the accuracy of the optical phase estimate is set by the Heisenberg's uncertainty principle \cite{GLM}. By contrast, the standard quantum limit (SQL) refers to the minimum level of quantum noise that can be obtained using standard approaches to phase estimation which do not involve real-time feedback. The SQL sets an important benchmark for the quality of a measurement and provides an interesting challenge to devise quantum strategies that can beat it.

Up until very recently, most of the work in quantum phase estimation has been on estimating a \emph{fixed} unknown phase shift. It was shown theoretically that \emph{adaptive} homodyne single-shot measurements can yield an estimate with mean-square error less than the SQL \cite{HMW,WK1,WK2}. This was subsequently demonstrated experimentally using very weak coherent states \cite{MA}. However, a more experimentally relevant problem is when the phase varies continuously under the influence of an unmeasured classical stochastic noise process \cite{BW,TSL,MT}.

The first experimental demonstration of adaptive quantum phase estimation of a continuously varying phase was presented in Ref. \cite{TW}, where an estimate could be obtained with a mean-square error of up to $2.24 \pm 0.14$ times smaller than the SQL. In the adaptive experiment, the system phase to be estimated was modulated by a classical stochastic Ornstein-Uhlenbeck (OU) noise process. We show here that the mean-square error in the estimate can be further reduced for the case of an OU noise process by using an optimal Kalman filter in the feedback loop and that the filter used in Ref. \cite{TW} is only optimal for the case where the noise process is a Wiener process and the measurement is assumed to be linear. See also Ref. \cite{PWL}.

It is physically unreasonable to specify precisely the desired values of the underlying parameters of the noise process modulating the system phase to be estimated. Hence it is desired to make the estimation process robust to uncertainty in these parameters. A robust quantum parameter estimation technique, as applied to atomic magnetometry, was demonstrated theoretically in Ref. \cite{JS}. Also, a simple approach to robust adaptive phase estimation was discussed briefly in Ref. \cite{PWL}. Robust adaptive estimation of the continuously varying optical system phase can be made by making the Kalman filter in the feedback loop robust to uncertainties introduced in the underlying parameters of the noise process. In this paper, this is achieved by using the guaranteed cost robust filtering approach described in Ref. \cite{PM}.

\section{MODEL OF ADAPTIVE PHASE ESTIMATION TECHNIQUE FROM REF. \cite{TW}}

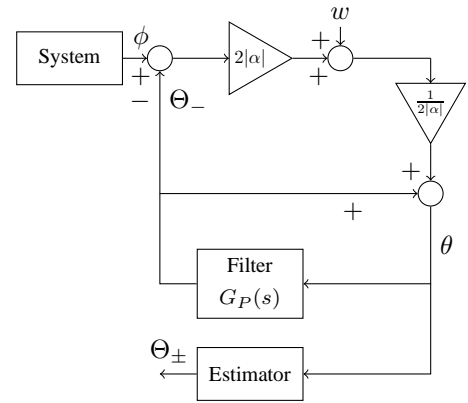
\begin{figure}[!b]
\centering
\begin{tikzpicture}[scale=0.6]
    \node [block] (system) at (0,0) {\footnotesize System};
    \node [sum] (sum) at (2,0) {};
    \node [pt] (gain) at (4,0) {$2|\alpha|$};
    \node [sum] (sum1) at (6,0) {};
    \node [open] (noise) at (6,1) {$w$};
    \node [none] (output) at (8,0) {};
    \node [pt, regular polygon rotate=180] (gain2) at (8,-1) {$\frac{1}{2|\alpha|}$};
    \node [none] (f) at (2,-3) {};
    \node [sum] (sum2) at (8,-3) {};
    \node [none] (g) at (8,-5) {};
    \node [block,text width=1cm,align=center] (filter) at (4,-5) {\footnotesize Filter\\ $G_P(s)$};
    \node [block] (estimator) at (4,-7) {\footnotesize Estimator};
    \node [none] (h) at (2,-7) {};

    \draw [->] (system) -- node [above, near end] {$\phi$} node [below, near end] {$+$} (sum);
    \draw [->] (sum) -- (gain);
    \draw [->] (gain) -- node [below, near end] {$+$} (sum1);
    \draw [->] (noise) -- node[left, near end] {$+$} (sum1);
    \draw [-] (sum1) -- (output);
    \draw [->] (output) -- (gain2);
    \draw [->] (gain2) -- node [left, near end] {$+$} (sum2);
    \draw [->] (sum2) -- node [right] {$\theta$} (g) -- (filter);
    \draw [->] (filter) -| (f) -- node [below, near end] {$+$} (sum2);
    \draw [->] (f) -- node [left, near end] {$-$} node [right, near end] {$\Theta_{-}$} (sum);
    \draw [->] (g) |- (estimator);
    \draw [->] (estimator) -- node [above, near end] {$\Theta_{\pm}$} (h);
\end{tikzpicture}
\caption{Block Diagram of the adaptive system considered in Ref. \cite{TW}.}
\label{fig:block_prl}
\end{figure}

Fig. \ref{fig:block_prl} shows the model block diagram of the adaptive system with filter used in Ref. \cite{TW} in the feedback loop. The \emph{estimator} calculates offline the estimate for the system phase.

\subsection{Process}

The governing equation for the adaptive estimation system being considered is the dynamically varying stochastic classical phase \cite{TW}:
\begin{equation}
d\phi(t) = - \lambda\phi(t)dt + \sqrt{\kappa}dV(t),
\end{equation}
where $\phi(t)$ is the system phase to be estimated, $dV$ is a Wiener increment, $\lambda > 0$ is the mean reversion rate and $\kappa > 0$ is the inverse coherence time.

The process model for the above system is, therefore, given by the equation:
\begin{equation}
\dot{\phi}(t) = -\lambda\phi(t) + \sqrt{\kappa}v(t),
\end{equation}
where $v := \frac{dV}{dt}$ is a white noise process with autocorrelation function $R_v(\tau) = \delta(\tau)$.

\subsection{Measurement}
Using a linearization approximation, the homodyne photocurrent from the adaptive phase estimation system is given by \cite{TW}:
\begin{equation}
I(t)dt = 2|\alpha|[\phi(t) - \hat{\phi}(t)]dt + dW(t),
\end{equation}
where $|\alpha|$ is the amplitude of the coherent state with photon flux given by $\mathcal{N} := |\alpha|^2$, $\hat{\phi}$ is the \emph{intermediate phase estimate} which is also the optical local oscillator phase because of the feedback, and $dW$ is Wiener noise arising from the quantum vacuum fluctuations.

The \emph{instantaneous estimate} $\theta(t)$ is given by:
\begin{equation}
\theta(t) := \hat{\phi}(t) + \frac{I(t)}{2|\alpha|}.
\end{equation}

The measurement model can, therefore, be described by the equation:
\begin{equation}
\theta(t) = \phi(t) + \frac{1}{2|\alpha|}w(t),
\end{equation}
where $w := \frac{dW}{dt}$ is also a white noise process with $R_w(\tau) = \delta(\tau)$.

\subsection{System Model}

Rewriting the equations for the system under consideration, we get
\begin{eqnarray}\label{eq:sys_model}
\boxed{
\begin{split}
\textsf{\small Process model:} \ \ \dot{\phi} &= -\lambda\phi + \sqrt{\kappa}v, \\
\textsf{\small Measurement model:} \ \ \theta &= \phi + \frac{1}{2|\alpha|}w,
\end{split}}
\end{eqnarray}
where
\begin{align*}
E[v(t)v(\tau)]& = \mathbf{Q}\delta(t - \tau), \\
E[w(t)w(\tau)]& = \mathbf{R}\delta(t - \tau), \\
E[v(t)w(\tau)]& = 0.
\end{align*}

Since $V$ and $W$ are Wiener processes, both $\mathbf{Q}$ and $\mathbf{R}$ are unity.

\subsection{Feedback}

\subsubsection{Transfer Function}

The feedback filter used for adaptive phase estimation in Ref. \cite{TW} has the following form:
\begin{equation}
\Theta_{-}(t) = \chi \int_{-\infty}^t \! \theta(\tau) e^{\chi.(\tau-t)} d\tau = \chi \left( \theta(t) * e^{-\chi .t} \right),
\end{equation}
where the value of $\chi$ is $\chi_{opt} = 2|\alpha|\sqrt{\kappa}$, which is optimal in the limit $\lambda \to 0$.

The transfer function of this filter can, thus, be found to be:
\begin{equation}\label{eq:tf_prl}
\boxed{G_P(s) = \frac{\Theta_{-}(s)}{\theta(s)} = \frac{\chi}{s + \chi}.}
\end{equation}

\subsubsection{Error Covariance}\label{sec:err_cov}

We augment the system given by (\ref{eq:sys_model}) with the feedback filter (\ref{eq:tf_prl}) as shown in Fig. \ref{fig:block_prl} and represent the augmented system by the state-space model:

\begin{equation}\label{eq:ss_model}
\mathbf{\dot{\overline{x}}} = \mathbf{\overline{A}\, \overline{x}} + \mathbf{\overline{B}\, \overline{w}},
\end{equation}
where

\qquad \qquad \( \mathbf{\overline{x}} = 
\left[ \begin{array}{c}
\phi \\
\Theta_{-}
\end{array} \right] \)
\qquad and \qquad
\( \mathbf{\overline{w}} = 
\left[ \begin{array}{c}
v \\
w
\end{array} \right]. \)

\vspace*{2mm}

From (\ref{eq:sys_model}) and (\ref{eq:tf_prl}), we obtain:
\begin{align}
\dot{\phi}& = -\lambda\phi + \sqrt{\kappa}v, \\
\dot{\Theta}_{-}& = - \chi_{opt}\Theta_{-} + \chi_{opt}\phi + \frac{\chi_{opt}}{2|\alpha|}w.
\end{align}

Thus, we have:

\( \mathbf{\overline{A}} = 
\left[ \begin{array}{cc}
-\lambda & 0 \\
\chi_{opt} & -\chi_{opt}
\end{array} \right] \)
\quad and \quad
\( \mathbf{\overline{B}} = 
\left[ \begin{array}{cc}
\sqrt{\kappa} & 0 \\
0 & \frac{\chi_{opt}}{2|\alpha|}
\end{array} \right]. \)

\vspace*{2mm}

For the continuous-time state-space model (\ref{eq:ss_model}), the steady-state state covariance matrix $\mathbf{P}_S$ is obtained by solving the \emph{Lyapunov equation}:
\begin{equation}\label{eq:lyapunov}
\mathbf{\overline{A}P}_S + \mathbf{P}_S\mathbf{\overline{A}}^T + \mathbf{\overline{B}\, \overline{B}}^T = 0,
\end{equation}
where $\mathbf{P}_S$ is the symmetric matrix
\[ 
\mathbf{P}_S = E(\mathbf{\overline{x}\, \overline{x}}^T) =
\left[ \begin{array}{cc}
P_1 & P_2 \\
P_2 & P_3
\end{array} \right].
\]

Upon solving (\ref{eq:lyapunov}), we get
\begin{align*}
P_1& = \frac{\kappa}{2\lambda},\\
P_2& = \frac{\chi_{opt}\kappa}{2\lambda(\lambda + \chi_{opt})},\\
P_3& = \frac{4\chi_{opt}|\alpha|^2\kappa + \chi_{opt}\lambda^2 + \chi_{opt}^2\lambda}{8|\alpha|^2\lambda(\lambda + \chi_{opt})}.
\end{align*}

The estimation error can be written as:
\[ \mathbf{e} = \phi - \Theta_{-} = [1 \, -1]\mathbf{\overline{x}}, \]
which is mean zero since all of the quantities determining $\mathbf{e}$ are mean zero.

The error covariance is then given as:

\begin{align*} 
\sigma_P^2& = E(\mathbf{ee}^T) = [1 \, -1]E(\mathbf{\overline{x}\, \overline{x}}^T)
\left[ \begin{array}{c}
1 \\
-1
\end{array} \right] \\ 
& =
[1 \, -1]
\left[ \begin{array}{cc}
P_1 & P_2 \\
P_2 & P_3
\end{array} \right]
\left[ \begin{array}{c}
1 \\
-1
\end{array} \right]
= P_1 - 2P_2 + P_3.
\end{align*}

Thus, we obtain:
\begin{equation}\label{eq:err_prl}
\boxed{\sigma_P^2 = \frac{\chi_{opt}(\lambda + 2\chi_{opt})}{8|\alpha|^2(\lambda + \chi_{opt})} = \frac{\sqrt{\kappa}(\lambda + 4|\alpha|\sqrt{\kappa})}{4|\alpha|(\lambda + 2|\alpha|\sqrt{\kappa})}.}
\end{equation}

One can verify that this analytical expression for the error covariance agrees with equation (10) of Ref. \cite{TW} for the optimal case of $\chi$.

\section{MODEL OF ADAPTIVE PHASE ESTIMATION USING A KALMAN FILTER}

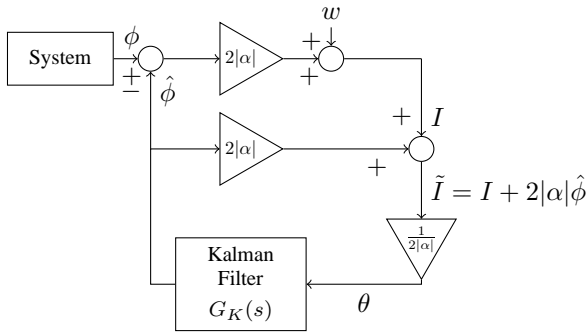
\begin{figure}[!b]
\centering
\begin{tikzpicture}[scale=0.6]
    \node [block] (system) at (0,0) {\footnotesize System};
    \node [sum] (sum) at (2,0) {};
    \node [pt] (gain) at (4,0) {$2|\alpha|$};
    \node [sum] (sum1) at (6,0) {};
    \node [open] (noise) at (6,1) {$w$};
    \node [none] (output) at (8,0) {};
    \node [none] (f) at (2,-2) {};
    \node [pt] (gain1) at (4,-2) {$2|\alpha|$};
    \node [sum] (sum2) at (8,-2) {};
    \node [pt, regular polygon rotate=180] (gain2) at (8,-4) {$\frac{1}{2|\alpha|}$};
    \node [block,text width=1.5cm,align=center] (kalman) at (4,-5) {\footnotesize Kalman Filter\\$G_K(s)$};

    \draw [->] (system) -- node [above, near end] {$\phi$} node [below, near end] {$+$} (sum);
    \draw [->] (sum) -- (gain);
    \draw [->] (gain) -- node [below, near end] {$+$} (sum1);
    \draw [->] (noise) -- node[left, near end] {$+$} (sum1);
    \draw [-] (sum1) -- (output);
    \draw [->] (output) -- node[left, near end] {$+$} node [right, near end] {$I$} (sum2);
    \draw [->] (gain1) -- node [below, near end] {$+$} (sum2);
    \draw [->] (sum2) -- node [right] {$\tilde{I}=I+2|\alpha|\hat{\phi}$}(gain2);
    \draw [->] (gain2) |- node [below, near end]  {$\theta$} (kalman);
    \draw [->] (kalman) -| (f) -- (gain1);
    \draw [->] (f) -- node [left, near end] {$-$} node [right, near end] {$\hat{\phi}$} (sum);
\end{tikzpicture}
\caption{Block Diagram of the adaptive system with a Kalman Filter in the feedback loop.}
\label{fig:block_kalman}
\end{figure}

Fig. \ref{fig:block_kalman} shows the block diagram of the adaptive system with a Kalman filter in the feedback loop. As compared to Fig. \ref{fig:block_prl}, there is a subtle difference in the way the input $\theta$ to the feedback filter is generated in this case.

\subsection{Algebraic Riccati Equation}

Given the system as described by (\ref{eq:sys_model}), the steady-state continuous Kalman filter Riccati equation \cite{RGB} can be shown to be:
\begin{equation}\label{eq:riccati}
\boxed{-2\lambda P - 4|\alpha |^2P^2 + \kappa = 0.}
\end{equation}

\subsection{Filter Equation}

The Kalman filter equation for the (continuous) system under consideration is then given by:
\begin{equation}\label{eq:kalman}
\boxed{\dot{\hat{\phi}} = - (\lambda + K)\hat{\phi} + K\theta,}
\end{equation}
where $K$ is the Kalman gain.

The stabilising solution of (\ref{eq:riccati}) can be found to be:
\begin{equation}\label{eq:errcov}
P = \frac{1}{4|\alpha|^2} \left( -\lambda + \sqrt{4\kappa|\alpha|^2 + \lambda^2} \right).
\end{equation}

The Kalman gain for the system under consideration is then given by:
\begin{equation}
K = -\lambda + \sqrt{4\kappa|\alpha|^2 + \lambda^2}.
\end{equation}

\subsection{Feedback}

\subsubsection{Transfer Function}

The transfer function of the Kalman filter may be obtained from (\ref{eq:kalman}) to be:
\begin{equation}\label{eq:tf_kalman}
\boxed{G_K(s) = \frac{\hat{\phi}(s)}{\theta(s)} = \frac{K}{s+\lambda+K}.}
\end{equation}

\subsubsection{Error Covariance}

The error covariance for the Kalman filter is given by (\ref{eq:errcov}), rewritten as below:
\begin{equation}\label{eq:err_kalman}
\boxed{\sigma_K^2 = \frac{K}{4|\alpha|^2} = \frac{1}{4|\alpha|^2} \left( -\lambda + \sqrt{4\kappa|\alpha|^2 + \lambda^2} \right).}
\end{equation}

One can verify that the error covariance for the Kalman filter, obtained using the Lyapunov method used in section \ref{sec:err_cov} earlier, would be the same as in the above equation.

\section{COMPARISON OF KALMAN FILTER WITH FILTER USED IN REF. \cite{TW}}

The filter used in the feedback loop for the adaptive system considered in Ref. \cite{TW} was designed to behave optimally under the condition that the value of $\lambda$ is zero, so as to approximate the detailed analysis in Ref. \cite{BW}. Here, we will see that a Kalman filter would be optimal in all cases, i.e. for all values of $\lambda$ and that the filter used in Ref. \cite{TW} coincides with the Kalman filter in the limit $\lambda \to 0$.

\subsection{Transfer Function}

Equations (\ref{eq:tf_prl}) and (\ref{eq:tf_kalman}) given earlier are the transfer functions of the filter used in Ref. \cite{TW} and the Kalman filter, respectively.

In the limit $\lambda \to 0$, we have
\[
G_K(s) \to \frac{K}{s + K},
\]
where
\[
K \to \lim_{\lambda \to 0} \left( -\lambda + \sqrt{4\kappa|\alpha|^2 + \lambda^2} \right) = 2|\alpha|\sqrt{\kappa} = \chi_{opt}.
\]

Let us consider the parameters for the adaptive phase estimation problem considered in Ref. \cite{TW}: $\mathcal{N}_{AP} = |\alpha|^2 = 1.3499 \times 10^6 s^{-1}$, $\kappa_{AP} = 1.5868 \times 10^4$ rad/s, $\lambda_{AP} = 6.1451 \times 10^4$ rad/s.

We can calculate the following, using the units as indicated above: $2|\alpha| = 2324$, $\sqrt{\kappa} = 126$, $\chi_{opt} = 292824$, and $K = 237643$.

Thus, we have:
\[ G_P(s) = \frac{292824}{s + 292824}, \]
whereas
\[ G_K(s) = \frac{237643}{s + 299094}. \]

The transfer function of the filter used in Ref. \cite{TW} is similar to, but not the same as, that of the optimal Kalman filter for the given set of parameters.

\subsection{Error Covariance}

Equations (\ref{eq:err_prl}) and (\ref{eq:err_kalman}) given earlier are the error covariances in the estimates from the filter used in Ref. \cite{TW} and the Kalman filter, respectively.

Note that in the limit $\lambda \to 0$, we have
\begin{equation}
\sigma_P^2 = \sigma_K^2 = \frac{\sqrt{\kappa}}{2|\alpha|}.
\end{equation}

Considering again the parameters for the adaptive phase estimation problem considered in Ref. \cite{TW}, we can evaluate:

\vspace*{2mm}

\( \qquad \sigma_P^2 = 0.0495, \)
\qquad whereas \qquad
\( \sigma_K^2 = 0.044. \)

\vspace*{2mm}

Thus, it is clear that the filter used in Ref. \cite{TW} is sub-optimal as compared to the use of a Kalman filter for the given value of $\lambda$.

However, in the limit $\lambda \to 0$, we would have:
\[
\sigma_P^2 = \sigma_K^2 = 0.0542.
\]

\begin{figure}[!b]
\hspace*{-8mm}
\includegraphics[width=0.57\textwidth]{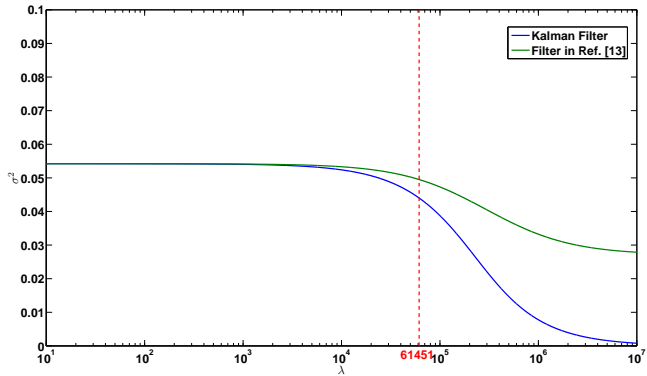}
\caption{Comparison of the error covariance between the Kalman filter and the filter used in Ref. \cite{TW}.}
\label{fig:comparison_graph}
\end{figure}

Fig. \ref{fig:comparison_graph} shows the plot of the error covariance against the parameter $\lambda$ for the two cases, viz. Kalman filter and the filter used in Ref. \cite{TW}. Here, we have used the nominal experimental values for the other two parameters, viz. $|\alpha|$ and $\kappa$, in the expressions (\ref{eq:err_prl}) and (\ref{eq:err_kalman}). As can be seen, the filter used in Ref. \cite{TW} behaves exactly like the optimal Kalman filter for lower values of $\lambda$. However, as the value of $\lambda$ rises, the Kalman filter's error covariance improves significantly as compared to that of the filter used in Ref. \cite{TW}. The red vertical line denotes the value of $\lambda$ for the adaptive experiment considered in Ref. \cite{TW}.

\section{ROBUST FILTER}

In this section, we make our filter robust to uncertainty in one of the underlying parameters using the guaranteed cost estimation robust filtering approach given in Ref. \cite{PM}.

\subsection{Process and Measurement}

We introduce uncertainty in the parameter $\lambda$ as follows:
\[ \lambda \rightarrow \lambda - \mu\lambda\Delta, \]
where $\Delta$ is an uncertain parameter which satisfies $|\Delta| \leq 1$ and $0 \leq \mu < 1$ is a parameter which determines the level of uncertainty in the model.

The process and measurement models of (\ref{eq:sys_model}) take the form:
\begin{eqnarray}\label{eq:robust_sys_model}
\boxed{
\begin{split}
\textsf{\small Process model:} \ \ \dot{\phi} &= (-\lambda + \mu\lambda\Delta)\phi + \sqrt{\kappa}v, \\
\textsf{\small Measurement model:} \ \ \theta &= \phi + \frac{1}{2|\alpha|}w.
\end{split}}
\end{eqnarray}

\subsection{Riccati Equation and Optimal Error Bound}

As in Ref. \cite{PM}, the Riccati equation for the guaranteed cost filter for the system is:
\begin{equation}\label{eq:riccati_robust}
\epsilon Q^2 - 4|\alpha|^2 Q^2 - 2\lambda Q + \frac{\mu^2 \lambda^2}{\epsilon} + \kappa = 0.
\end{equation}

The stabilising solution of this equation yields an upper bound for the robust filter error covariance:
\small
\begin{equation}
Q^+ = \frac{-\lambda \epsilon + \sqrt{\lambda^2\epsilon^2 - \epsilon^3\kappa - \epsilon^2\mu^2\lambda^2 + 4|\alpha|^2\epsilon^2\kappa + 4|\alpha|^2\epsilon\mu^2\lambda^2}}{\epsilon(-\epsilon + 4|\alpha|^2)}.
\end{equation}
\normalsize

The optimum value of $\epsilon$ at which the bound $Q^+$ is the minimum can be found to be:
\small
\begin{equation}
\epsilon_{opt} = \frac{\left(-\lambda\mu + \lambda + \sqrt{\mu^2\lambda^2 - 2\lambda^2\mu + \lambda^2 + 4|\alpha|^2\kappa}\right)\mu\lambda}{\kappa}.
\end{equation}
\normalsize

\subsection{Filter Equation}

We calculate the robust filter equation for our system to be \cite{PM}:
\small
\begin{equation}\label{eq:robust}
\boxed{\dot{\hat{\phi}} = -\lambda\hat{\phi} + (\epsilon - 4|\alpha|^2)Q^+\hat{\phi} + 4|\alpha|^2Q^+\phi + 2|\alpha|Q^+w.}
\end{equation}
\normalsize

Note that when $\mu = 0$, then $\epsilon_{opt} = 0$ and (\ref{eq:riccati_robust}) reduces to (\ref{eq:riccati}) with $P$ replaced by $Q$. That is, the robust filter reduces to the standard Kalman filter.

\subsection{Transfer Function}

Equation (\ref{eq:robust}) yields the below transfer function for the robust filter:

\begin{equation}\label{eq:tf_robust}
\boxed{G_R(s) = \frac{\hat{\phi}(s)}{\theta(s)} = \frac{4|\alpha|^2Q^+}{s+\lambda+(4|\alpha|^2-\epsilon)Q^+}.}
\end{equation}

This transfer function is a first-order low-pass filter with gain and corner frequency slightly different from those of the filter used in Ref. \cite{TW} and the Kalman filter.

\section{COMPARISON OF ROBUST FILTER WITH KALMAN FILTER}

We can compute using the Lyapunov method employed in section \ref{sec:err_cov} and for the nominal experimental values of all the parameters and $\mu = 0.05$, the error covariance for our robust filter as a function of $\Delta$. We can similarly compute the error covariance as a function of $\Delta$ for the Kalman filter (\ref{eq:kalman}) where the process and measurement models are for the uncertain system given by (\ref{eq:robust_sys_model}).

\begin{figure}[!t]
\hspace*{-5mm}
\includegraphics[width=0.57\textwidth]{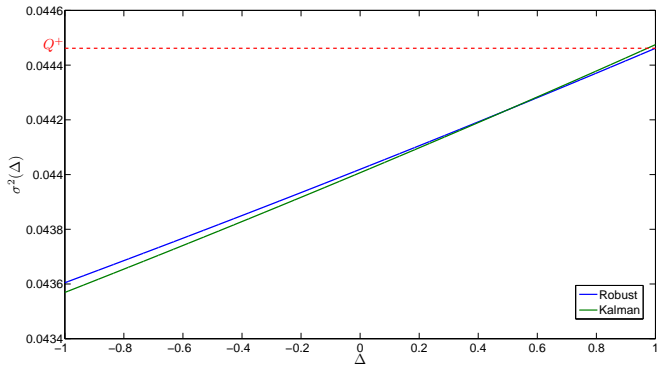}
\caption{Comparison of error covariance as a function of $\Delta$ for $\mu = 0.05$.}
\label{fig:robust_vs_kalman1}
\end{figure}

We can then obtain a plot of the error covariance as a function of $\Delta$ for the robust filter and the Kalman filter on the same graph, as shown in Fig. \ref{fig:robust_vs_kalman1}. As we can see from this figure, there is not a significant performance improvement with the robust filter as compared to the Kalman filter for the case of $5\%$ uncertainty.  The horizontal dashed line indicates the optimal error bound for the robust filter.

\begin{figure}[!htb]
\hspace*{-5mm}
\includegraphics[width=0.57\textwidth]{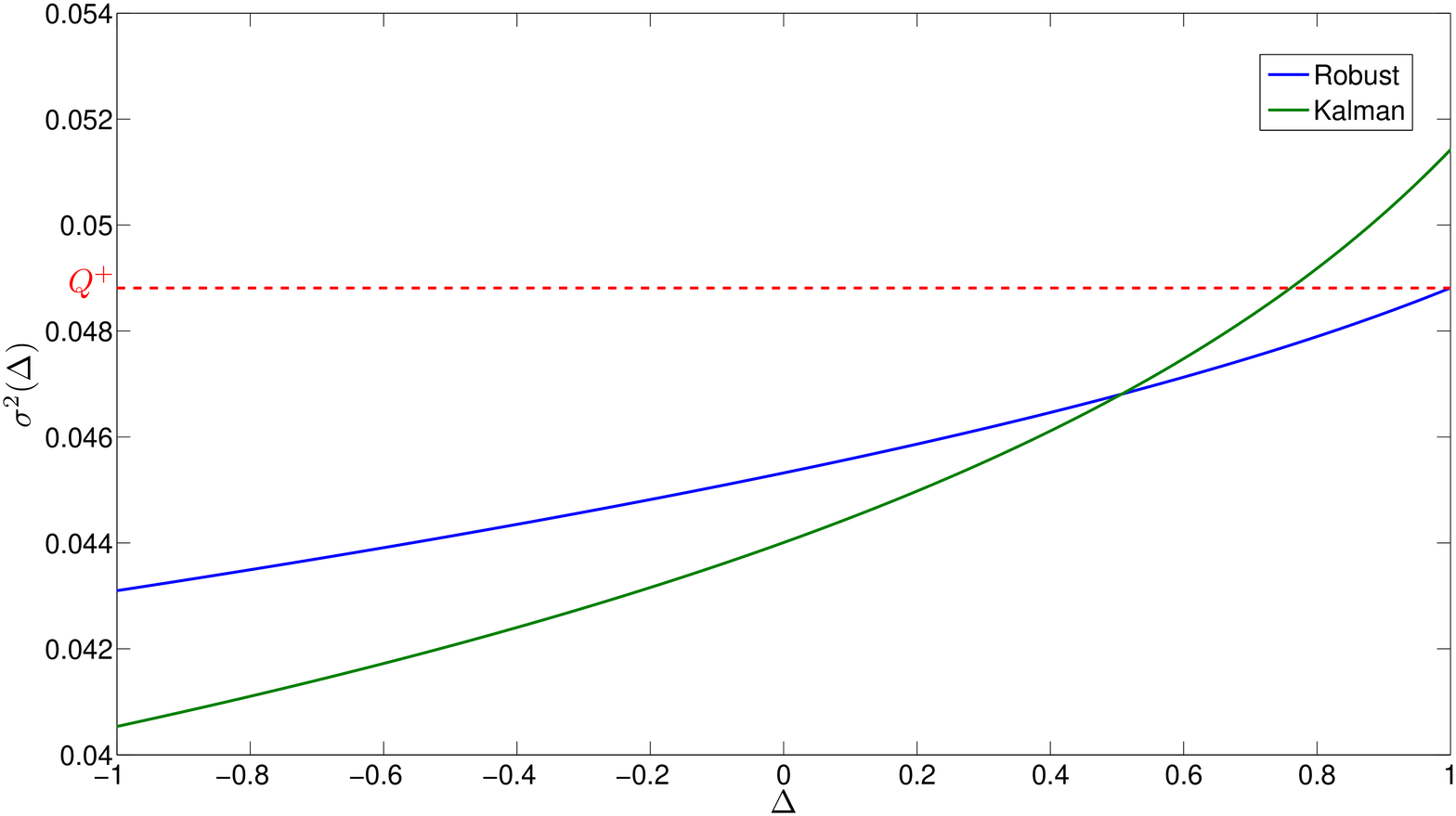}
\caption{Comparison of error covariance as a function of $\Delta$ for $\mu = 0.5$.}
\label{fig:robust_vs_kalman2}
\end{figure}

\begin{figure}[!htb]
\hspace*{-5mm}
\includegraphics[width=0.57\textwidth]{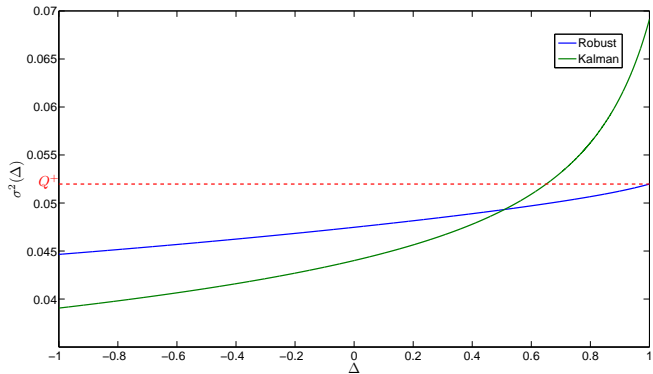}
\caption{Comparison of error covariance as a function of $\Delta$ for $\mu = 0.8$.}
\label{fig:robust_vs_kalman3}
\end{figure}

We can, similarly, plot such comparison graphs for the case of more uncertainty in $\lambda$. Figs. \ref{fig:robust_vs_kalman2} and \ref{fig:robust_vs_kalman3} show plots for $\mu = 0.5$ and $\mu = 0.8$, respectively. As we can see, the robust filter performs better than the Kalman filter as $\Delta$ approaches $1$ for all levels of uncertainty in $\lambda$. As the uncertainty in $\lambda$ increases, so does the improvement in performance of the robust filter relative to the Kalman filter.

\section{CONCLUSION}

This paper applies the theory of robust filtering to the problem of adaptive homodyne estimation of a continuously evolving optical phase shift of a quantum system. The immediate further work to this would be to extend the theory to include Kalman and robust smoothing rather than filtering alone. It remains to illustrate the results in this article experimentally. The results herein may as well be extended for the case of squeezed states of light. Also, it would be interesting to explore robustness as applied to other types of complex noise processes or uncertainties in other parameters such as the noise power or photon flux.

\addtolength{\textheight}{-12cm}   % This command serves to balance the column lengths
                                  % on the last page of the document manually. It shortens
                                  % the textheight of the last page by a suitable amount.
                                  % This command does not take effect until the next page
                                  % so it should come on the page before the last. Make
                                  % sure that you do not shorten the textheight too much.

%%%%%%%%%%%%%%%%%%%%%%%%%%%%%%%%%%%%%%%%%%%%%%%%%%%%%%%%%%%%%%%%%%%%%%%%%%%%%%%%

%%%%%%%%%%%%%%%%%%%%%%%%%%%%%%%%%%%%%%%%%%%%%%%%%%%%%%%%%%%%%%%%%%%%%%%%%%%%%%%%

%%%%%%%%%%%%%%%%%%%%%%%%%%%%%%%%%%%%%%%%%%%%%%%%%%%%%%%%%%%%%%%%%%%%%%%%%%%%%%%%

\bibliographystyle{IEEEtran}
\bibliography{IEEEabrv,robustbiblio}

\end{document}